# Time- and angle-resolved photoelectron spectroscopy of strong-field light-dressed solids: prevalence of the adiabatic band picture

Ofer Neufeld[1,*], Wenwen Mao[1], Hannes Hübener[1], Nicolas Tancogne-Dejean[1], Shunsuke A. Sato[1,2], Umberto De Giovannini[1,3], Angel Rubio[1,4,*]

[1]Max Planck Institute for the Structure and Dynamics of Matter and Center for Free-electron Laser Science, Hamburg 22761, Germany.

[2]Center for Computational Sciences, University of Tsukuba, Tsukuba 305-8577, Japan.

[3]Università degli Studi di Palermo, Dipartimento di Fisica e Chimica—Emilio Segrè, Palermo I-90123, Italy.

[4]Center for Computational Quantum Physics (CCQ), The Flatiron Institute, New York 10010, USA.

*Corresponding author E-mails: oneufeld@schmidtsciencefellows.org, angel.rubio@mpsd.mpg.de.

**Abstract**

In recent years, strong-field physics in condensed-matter was pioneered as a novel approach for controlling material properties through laser-dressing, as well as for ultrafast spectroscopy *via* nonlinear light-matter interactions (e.g. harmonic generation). A potential controversy arising from these advancements is that it is sometimes vague which band-picture should be used to interpret strong-field experiments: the field-free bands, the adiabatic (instantaneous) field-dressed bands, Floquet bands, or some other intermediate picture. We here try to resolve this issue by performing 'theoretical experiments' of time- and angle-resolved photoelectron spectroscopy (Tr-ARPES) for a strong-field laser-pumped solid, which should give access to the actual observable bands of the irradiated material. To our surprise, we find that the adiabatic band-picture survives quite well, up to high field intensities ($\sim 10^{12}$ W/cm$^2$), and in a wide frequency range (driving wavelengths of 4000 to 800nm, with Keldysh parameters ranging up to $\sim 7$). We conclude that to first order, the adiabatic instantaneous bands should be the standard blueprint for interpreting ultrafast electron dynamics in solids when the field is highly off-resonant with characteristic energy scales of the material. We then discuss weaker effects of modifications of the bands beyond this picture that are non-adiabatic, showing that by using bi-chromatic fields the deviations from the standard picture can be probed with enhanced sensitivity. Our work outlines a clear band picture for the physics of strong-field interactions in solids, which should be useful for designing and analyzing strong-field experimental observables and also to formulate simpler semi-empirical models.

# I. INTRODUCTION

When solids are irradiated by strong laser pulses, a plethora of physical effects can be initiated. For instance, these can include high harmonic generation (HHG) [1,2], creation of transient injection currents [3–8], the dynamical Franz-Keldysh effect [9,10], creation of new laser-dressed states of matter [11–13], and more. Depending on the symmetries of the solid and the properties of the laser field, one can even create transient topological phases [12,14], or valley-polarization by selectively populating certain regions of the Brillouin zone [13,15–18]. These phenomena are of technological importance for quantum information processing and petahertz electronics [19], since they allow to generate, and probe, ultrafast motion of electrons in solids, in real-time. Of particular interest are topological [20–27] and strongly-correlated [28–32] phases, since these could hold the key for novel applications.

What is common in these effects is that they all occur under the same electronic conditions – strong-field driving of a periodic system. However, each effect is usually analyzed in terms of a different band picture to rationalize the results and physical mechanisms. For instance, the mechanism of HHG is usually described with semi-classical models employing the field-free bands [33–42], which are utilized for HHG-spectroscopy of various basic electronic properties such as band structures berry-curvatures and density of states [43–47]. The dynamical Franz-Keldysh effect can be interpreted on the other hand with a Floquet time-averaged picture for the field-dressed bands [9,10,48–50]. For longer wavelength driving, one might employ an adiabatic band picture based on the instantaneous states of the laser-dressed system, the so called Houston states [51,52], which is useful to obtain insight about vector-potential induced shifting of the bands [53]. Lastly, some phenomena require an intermediate picture that is neither fully time-averaged in the Floquet sense, nor instantaneous like the Houston states, which was recently utilized for measuring transient gap closings in higher conduction bands [54]. On the other hand, there are works demonstrating that in the same conditions, there can be very strong modification of the band structure due to the laser driving itself. For resonant interactions in the weak-field regime, this approach is known to yield Floquet-topological insulators [14]. But naturally, strong-field driving can also yield similar topological phases where the system's gap substantially changes [13], or even other phases such as free-electron like states [12,55,56]. In these cases, one might say that the choice of how to interpret experiments and formulate models for the electron dynamics is somewhat vague – after all, even if any full basis set can still be used for formulating a model, there is no point in describing electron trajectories in bands if those bands do not exist or alter strongly over time. Similarly, it might be confusing to use dressed-bands for describing new phases of matter if the underlying electron dynamics are well-captured by the field-free bands. We emphasize that the complication arises since several of the effects can co-exist because they are driven in the same conditions (they are just described by different observables). Beyond this, it is still an open question up to what laser powers, driving wavelengths, and Keldysh parameters, the band picture holds – i.e. do the band shapes significantly change in the strong-field regime (but still below the material damage threshold)?

One possible solution is to employ time- and angel-resolved photoelectron spectroscopy (Tr-ARPES) [57–59]. This approach allows to directly image populated single-particle bands in periodic systems even if they are simultaneously interacting with strong lasers. The advantage here is that the bands themselves become the observable quantities, rather than the basis set used for interpreting the experiment. ARPES has already been successfully employed for probing the electronic structure of many systems [60], including topological insulators and Weyl semimetals [61], superconductors [62], and valley pseudospin [63,64]. It has also been utilized for probing dynamical phenomena such as photocurrents in topological insulators [65] and electron-phonon couplings [66–68]. However, it has not yet been employed for investigating a solid system dressed by a strong laser field, which might provide new insights.

Here we perform *ab-initio* calculations of 'numerical experiments' of Tr-ARPES in a periodic system that is dressed by intense laser light. Specifically, we explore a monolayer of hexagonal-boron-nitride (hBN) as a benchmark system, and investigate an instantaneous regime (that probes the instantaneous field-dressed bands), and a Floquet regime (that probes the Floquet cycle-averaged bands). We find that the instantaneous band picture holds up extremely well in strong-field driving – there is no observable breakdown of the bands or strong modification of the energy scales even in conditions where harmonics up to ~25 eV are driven (driving powers of ~$10^{12}$ W/cm$^2$), and in wavelengths ranges of 4000 to 800nm (Keldysh parameters <7). That is to say that the band picture persists, even far away from what is standardly considered as the adiabatic regime with Keldysh parameters <1. The main effect of the driving is seen to tune the phases of the electronic states (which affects transition matrix elements), and to shift the band origin with the instantaneous vector potential (as explained by the adiabatic theory) [51,52]. We also find that some modification of the bands arise beyond this instantaneous picture, i.e. involving electron dynamics that are non-adiabatic. These are slightly weaker effects, but should be experimentally detectable. We further show that by using a bi-chromatic dressing field, the non-adiabaticity can be probed with enhanced sensitivity.

The paper is ordered as follows: In Section II we present the methodology of our *ab-initio* calculations, also detailing how to remove contributions of continuum electron acceleration from the analysis. Section III outlines the logic of 'numerical experiments' of Tr-ARPES from strongly driven solids, outlining different regimes of laser parameters, as well as some of the expected band pictures as predicted from various theories. Section IV presents the main results and analysis, showing field-dressed bands under various driving conditions, including both Floquet and instantaneous regimes, and discussing adiabatic and non-adiabatic effects. Section V presents Tr-ARPES from bi-chromatically strong-field driven systems with emphasis on non-adiabatic light-dressing effects. Finally, section VI summarizes our results.

## II. METHODOLOGY

### A. Laser-driven dynamics

We begin by outlining the methodology of our *ab-initio* calculations. All results are obtained with the open-access real-space grid-based code, Octopus [69–72]. The dynamics of electrons with impinging laser light are described within the framework of real-time time-dependent density functional theory (TDDFT) [73], and within the adiabatic approximation for the exchange correlation (XC) functional (in the local density approximation). The laser-matter interaction is described in the velocity gauge, where the ions are assumed to be frozen and non-interacting with the laser field (which should be a valid approximation on few femtosecond timescales). We also neglect any macroscopic effects of phase matching and laser pulse propagation in the material. The resulting equations of motion for the Kohn-Sham (KS) Bloch states are given by (we use atomic units throughout, unless stated otherwise):

$$i\partial_t |\varphi_{n,k}^{KS}(t)\rangle = \left(\frac{1}{2}\left(-i\nabla + \frac{\mathbf{A}(t)}{c}\right)^2 + v_{KS}(\mathbf{r},t)\right)|\varphi_{n,k}^{KS}(t)\rangle \tag{1}$$

where $|\varphi_{n,k}^{KS}(t)\rangle$ is the KS-Bloch state at *k*-point *k* and band index *n*, $\mathbf{A}(t)$ is the total vector potential of all laser pulses interacting with matter within the dipole approximation, such that $-\partial_t \mathbf{A}(t) = c\mathbf{E}(t)$, *c* is the speed of light in atomic units ($c \approx 137.036$). $v_{KS}(\mathbf{r},t)$ in Eq. (1) is the time-dependent KS potential given by:

$$v_{KS}(\mathbf{r},t) = -\sum_I \frac{Z_I}{|\mathbf{R}_I - \mathbf{r}|} + \int d^3r' \frac{n(\mathbf{r}',t)}{|\mathbf{r}-\mathbf{r}'|} + v_{XC}[n(\mathbf{r},t)] \tag{2}$$

where $Z_I$ is the charge of the *I*'th nuclei and $\mathbf{R}_I$ is its coordinate, $v_{XC}$ is the XC potential that is a functional of $n(\mathbf{r},t) = \sum_{n,k}|\langle\mathbf{r}|\varphi_{n,k}^{KS}(t)\rangle|^2$, the time-dependent electron density. We note that practically, the bare Coulomb interaction of electrons with the nuclei is replaced with non-local pseudopotentials to reduce numerical costs (assuming the frozen core approximation for deep atomic states).

Following Eqs. (1) and (2), the KS Bloch wave functions are propagated in time, where the initial states are taken as the ground state which is found within DFT. All technical details of the propagation scheme and DFT calculations are delegated to the appendix.

### B. ARPES calculations

In order to obtain ARPES spectra, we follow the path of performing 'numerical experiments', which effectively simulate the measurement process as it would occur in the lab (to the best currently possible extent). In this respect, the total vector potential in Eq. (1) that is used for propagating the KS states is split to two:

$$\mathbf{A}(t) = \mathbf{A}_{\text{pump}}(t) + \mathbf{A}_{\text{probe}}(t) \tag{3}$$

Here $\mathbf{A}_{\text{pump}}(t)$ refers to the strong-laser field that dresses the system. This field is taken to have the form:

$$\mathbf{A}_{\text{pump}}(t) = f(t)\frac{cE_0}{\omega}\sin(\omega t)\hat{\mathbf{e}} \tag{4}$$

where $f(t)$ is an envelope function (see appendix for details), $E_0$ is the field amplitude, $\omega$ is the carrier frequency, and $\hat{\mathbf{e}}$ is a unit vector pointing transversely to the B-N bond axis (along the Γ-K line in reciprocal space) within the monolayer plane (*xy* plane). The probe pulse, $\mathbf{A}_{\text{probe}}(t)$, is taken as a linearly-polarized pulse polarized along the *z*-axis (transverse to the monolayer), with an extreme ultraviolet (XUV) carrier frequency:

$$\mathbf{A}_{\text{probe}}(t) = f_{\text{xuv}}(t-t_0)\frac{cE_{\text{xuv}}}{\omega_{\text{xuv}}}\sin(\omega_{\text{xuv}}t)\hat{\mathbf{z}} \tag{5}$$

where $f_{\text{xuv}}(t)$ is an envelope function with a maximum at $t=0$, $t_0$ indicates the pump-probe delay, $\omega_{\text{xuv}}$ is the probe photon frequency taken here at 100eV in order to obtain high temporal resolution, and $E_{\text{xuv}}$ is the probe's amplitude which is taken in the weak-field linear-response regime, corresponding to an intensity of $10^8$ W/cm$^2$. Thus, the pump laser field in Eq. (4) effectively dresses the solid, potentially causing some modification of the band structure, and the probe in Eq. (5) describes an XUV isolated pulse that photoionizes electrons which are then recorded in a momentum-resolved manner as the ARPES spectra. We note that the pump-probe delay, $t_0$, will be referred to from this point on as $t_{ion}$, which indicates a particular moment in time where $f_{\text{xuv}}$ is maximized (leading to the highest probability for photoemission) with respect the $\mathbf{A}_{\text{pump}}(t)$. For example, the notation $\mathbf{A}_{\text{pump}}(t_{ion})=0$ indicates that $t_0$ is set such that $f_{\text{xuv}}$ is maximized at the moment in time when $\mathbf{A}_{\text{pump}}=0$, but closest to the center of the pump pulse (i.e. closest to where $f(t)$ is maximized). Fig. 1(a) summarizes the schematic of the set-up.

The ARPES spectra are calculated directly from the propagated KS states, and without additional fundamental assumptions, using the highly accurate and efficient surface-flux method T-SURFF [74,75]. Here the momentum-resolved flux of photoelectrons is recorded across a surface normal to the hBN monolayer (located at $z=0$), which lies sufficiently far away in the continuum to avoid interactions of outgoing waves with the monolayer (the surface is located at the plane $z=w$). The photoemission from all KS Bloch states is coherently summed, producing *k*-resolved spectra along high-symmetry axes. For hBN, we calculate ARPES along a path in reciprocal space that traverses from Γ to K and K' (which is also taken as the pump laser driving axis in Eq. (4)). An exemplary ARPES spectra for the field-free case (with $E_0=0$) can be seen in Fig. 1(c), side-by-side with the field-free band-structure obtained with DFT (Fig. 1(b)). Additional technical details can be found in the appendix.

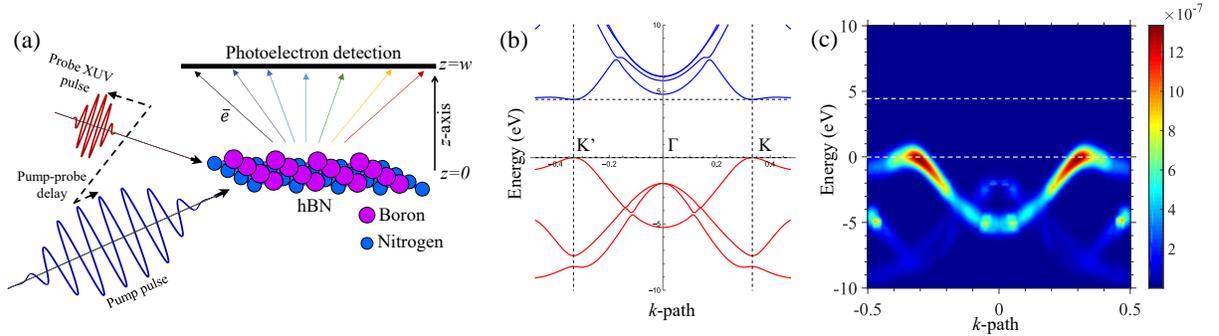

FIG. 1. (a) Illustration of Tr-ARPES with strong-field set-up. A monolayer of hBN is irradiated by an intense pump pulse that is polarized in the monolayer plane (blue), which dresses the electronic system and potentially causes some band modifications. A delayed XUV probe pulse that is polarized transverse to the monolayer (along the *z*-axis) then photoionizes electrons into the continuum. The photoemitted electrons propagate up to the detector surface, at which point their momenta and energy is recorded. (b) Ground-state DFT band structure along K'-Γ-K path. The top of the valence band is set to zero energy, and K and K' points are indicated with dashed lines. Occupied bands are denoted by red lines, while unoccupied bands with blue lines. (c) Pump-free ARPES spectra that captures the details of the equilibrium occupied bands. The energy of the ARPES spectrum is offset by ω$_{\text{xuv}}$, and dashed white lines indicate the positions of the valence and conduction band edges.

## C. Removing continuum effects

One issue with the approach (and corresponding experimental set-up) above, is that substantial modifications to the Tr-ARPES spectra might arise from propagation of the photoelectrons in the continuum up to the detector (e.g., laser-assisted photoemission [76,77]), rather than from the light-driven states inside matter. Usually, these effects comprise a main physical result (e.g. in photoemission from atomic and molecular species [78]). However, here we are interested in uncovering the nature of the field-dressed bands in the solid, and continuum-related phenomena can be considered as spurious effects that should be artificially removed. This can be a cumbersome procedure, because it essentially requires that $\mathbf{A}_{\text{pump}}(t)$ should only be 'active' in a certain region of the simulation box that is localized around the monolayer. One potential solution could be to multiply $\mathbf{A}_{\text{pump}}(t)$ by a spatial envelope, but this is not applicable in the dipole approximation and presents additional complications.

Therefore, we employ two other procedures in order to remove continuum effects from ARPES spectra. First, the photoelectrons that reach the 'numerical detector' (the surface through which the flux is calculated) are not further propagated in time, but rather, their momenta are recorded at that exact moment. This is a different approach than that usually employed in surface-flux methods, where the photoelectrons are further propagated under the full vector potential as Volkov states (assuming a strong-field approximation) [74,79]. This procedure guarantees that continuum-related effects correspond only to the laser-driven acceleration in the spatial region spanning from the monolayer up to the detector, which should be small because that region is only 30 Bohr wide. An exemplary result of this approach can be seen in Fig. 2(a,b) that compares field-dressed Tr-ARPES spectra with/without propagation of the Volkov states. From comparing Fig. 2(a) and Fig. 2(b), it is clear that propagation of electrons in the continuum causes substantial modifications of the ARPES spectra. These modifications will be discussed in more detail in the next subsection.

Notably, because $E_0$ in $\mathbf{A}_{\text{pump}}$ is so large in the strong-field regime, continuum-related effects can still arise after performing the procedure described above – photoelectrons can still accelerate in the *xy* plane (because they interact with $\mathbf{A}_{\text{pump}}$) during their propagation up to the detector. For instance, one still observes some linear sloping of the bands in Fig. 2(b), which could partially arise from continuum acceleration rather than from laser-dressing of the solid (because laser-induced acceleration adds a term linearly proportional to **k** in the photoelectron energy, see below). Therefore, we employ a second procedure to remove such effects – we assume that in the region spanning from the monolayer up to the detector, the electrons propagate semi-classically, such that their overall momentum is varied by **ΔA**:

$$\Delta \mathbf{A}(t_{ion}) = -c \int_{t_{ion}}^{t_{ion}+\Delta t} \mathbf{E}_{\text{pump}}(t) dt \tag{6}$$

where $t_{ion}$ indicates the approximate moment of ionization (taken here as the time that maximizes $\mathbf{A}_{\text{probe}}(t)$, assuming instantaneous ionization), $\Delta t$ corresponds to the semi-classical time-of-flight of the photoelectron as it traverses from $z = 0$ at the monolayer, up to the detector at $z = w$, and $\mathbf{E}_{\text{pump}}(t)$ is the electric field associated with $\mathbf{A}_{\text{pump}}(t)$, $c\mathbf{E}_{\text{pump}}(t) = -\partial_t \mathbf{A}_{\text{pump}}(t)$. $\Delta t$ is calculated assuming semi-classical dynamics under the strong-field approximation – the velocity component of the outgoing electron along the *z*-axis is taken as a constant of motion:

$$v_z = \sqrt{2\omega_{\text{xuv}}} \tag{7}$$

, such that $\Delta t = w/v_z$. Following this, the ARPES spectra can be artificially subtracted with the energy term $\mathbf{k} \cdot \Delta \mathbf{A}(t_{ion})/c$, were **k** denotes the crystal momentum. The result of this procedure can be seen in Fig. 2(c), which further corrects the main linear slope in the bands seen in Fig. 2(b).

Importantly, the second procedure that subtracts propagation effects in the simulation box can only be performed if a singular well-defined moment of ionization with respect to $\mathbf{A}_{\text{pump}}(t)$ exists. This only occurs in a regime where the probe pulse duration is much shorter than a laser cycle (see below). Thus, the second correction is not applied to the Floquet-regime calculations outlined below (where the probe pulse duration is longer than a laser cycle). We also emphasize that it is still an approximate procedure, because it neglects interactions of the outgoing electron with the monolayer, with the other electrons, and assumes semi-classical motion. Moreover, we have assumed an instantaneous moment of photoionization and neglected possible time-delays [80,81]. Still, errors arising from this approximation are expected to be small, because: (i) $\omega_{xuv}$ is taken to be very large (100 eV), which minimizes the time-of-flight of photoelectrons to ~200 attoseconds, and (ii) such effects are anyways usually quite small.

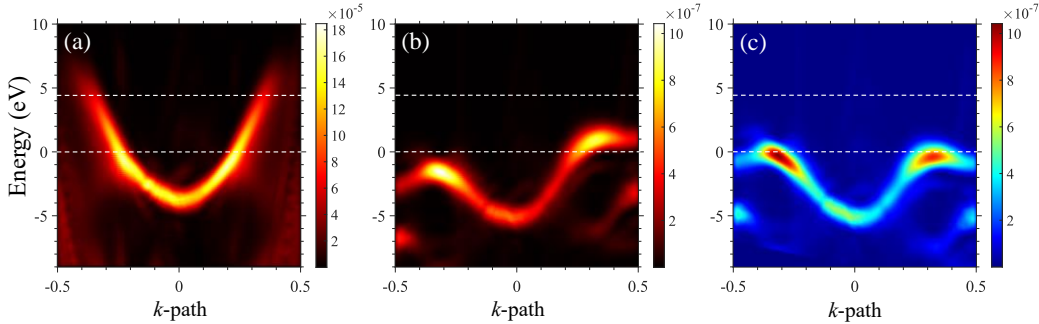

FIG. 2. Continuum effects in strong-field dressed Tr-ARPES at a pump-probe delay that corresponds to $\mathbf{A}(t_{ion})$=0. (a) Tr-ARPES spectrum from hBN including full propagation in the continuum. (b) Tr-ARPES spectrum from hBN including only propagation in the continuum within the simulation box, from $z=0$ to $z=w$. The propagation beyond $z=w$ is not performed (see text). (c) Tr-ARPES spectrum from hBN after removing continuum effects according to the two procedures outlined in the text. All Tr-ARPES spectra are calculated in the instantaneous regime (see text) with the parameters $I_0=10^{12}$ W/cm$^2$, $\lambda$=2000 nm. A different color scale is used for the case where all continuum effects have been removed to indicate that it represents the approximate energy spectrum of the electrons inside the dressed solid. Dashed white lines indicate the positions of the valence and conduction band edges.

### D. Continuum effects

At this point, we briefly discuss some of the continuum-related effects that can occur in the calculation of Tr-ARPES involving strong fields. This discussion is especially important in light of the context for our work – if one does not take extra care in removing these effects, it may seem as though unique new phases of matter are measured, while modifications of the bands could just result from continuum-dressing.

As an example, we highlight some continuum effects in Fig. 2: Figure 2(a) presents Tr-ARPES spectra in the standard conditions, where outgoing waves include all of the continuum effects, and the pump-probe delay is set such that $\mathbf{A}(t_{ion}) = 0$. For this pump-probe delay, the instantaneous adiabatic band picture corresponds to field-free system (because $\mathbf{A} \cong 0$ at the moment when the photoelectron is ionized), such that the predicted Houston states coincide with the field-free states. However, as clearly visible, the Tr-ARPES spectra shows a parabolic band dispersion along the pump laser polarization axis. The band energies strongly deviate from the field-free system, especially around the K and K' points at the valence band maxima (VBM), seemingly predicting a closing of the band gap. If one were to interpret this 'measurement' without post-processing at face value, it would suggest that the strong-field induces a free-electron-like state inside the material. However, as is obvious from Fig. 2(b), this free-electron-like state is only present in the dressed continuum. Moreover, deeper-lying bands also exhibit this behavior (not shown), clearly indicating that the parabolic bands originate from continuum acceleration of photoelectrons and not from a laser modification of the solid, which would not affect all the bands equally. Similarly, other phenomena may appear for other pump-probe delays, which could be perceived as fingerprints for unique light-driven phases, such as band bending or widening. We stress that extra caution must be taken before analyzing such phenomena, as they might result from continuum-based effects, even after the large majority of continuum effects have been artificially removed. For instance, some band widening along the energy axis is expected due to the non-singular ionization times of photoelectrons. That is, because electrons can also be ionized slightly before or after the assumed $t_{ion}$, variations in $\mathbf{\Delta A}$ can be incurred as the outgoing waves propagate to the detector, which cannot be fully subtracted out of the ARPES spectra.

In an effort to clarify some of these effects, all ARPES spectra in this paper that present the physical field-dressed bands are represented in blue-red colormaps. ARPES spectra that include some continuum effects that have not been subtracted with the two procedures described above are represented by orange-black colormaps. We emphasize that some weaker continuum-related effects might be present in the ARPES spectra that have undergone the continuum-removal procedures.

### III. DRESSED-BAND PICTURES AND LASER REGIMES

Having outlined the methodological approach, we now detail the different possible electronic band pictures for interpreting the dynamics, and the physical regimes of interest. We also analyze the main degrees of freedom in $\mathbf{A_{pump}}(t)$ and $\mathbf{A_{probe}}(t)$ and their effect on the resulting spectra.

First, we outline two regimes in which Tr-ARPES is employed in this work. The first regime is denoted as the 'instantaneous regime'. Here, the probe pulse is very short compared to the laser cycle of $\mathbf{A_{pump}}(t)$. Practically, this means that we have:

$$FWHM_{xuv} \ll 2\pi/\omega \tag{8}$$

where $FWHM_{xuv}$ is the full-width-half-max of the probe pulse. An illustration for this regime can be seen in Fig. 3(a) – the probe pulse samples a particular time in the pumped system. The resulting ARPES spectra should then roughly correspond to the band-structure of the system in that moment, whatever that may be. The main idea is that in this way one may probe the actual physical states in the system as it interacts with $\mathbf{A_{pump}}(t)$, and resolve any ambiguity that arises from the particular theory one uses to define the state of the system.

The second regime is denoted as the 'Floquet regime'. In this regime the probe pulse duration is on a similar order of magnitude as (or larger than) the pump laser cycle:

$$FWHM_{xuv} \approx 2\pi/\omega \tag{9}$$

Here one cannot assign a specific moment of ionization to photoelectrons with respect to the laser cycle, but rather, electrons are emitted continuously throughout different times along the pump field. The photoemission from all of these moments is summed coherently, yielding a type of cycle-averaged band picture, which should correspond to the Floquet limit of the field-dressed system (see Fig. 3(b) for illustration).

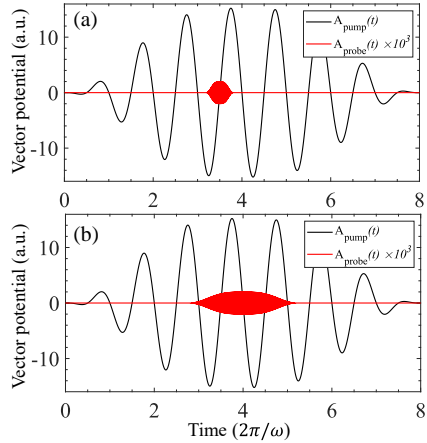

FIG. 3. Illustration of the Tr-ARPES regimes. (a) Instantaneous regime: the probe pulse has a duration much shorter than the pump field's optical cycle. In this exemplary case, the pump-probe delay is set such that the probe overlaps with a momentary zero in the vector potential, i.e. $\mathbf{A_{pump}}(t_{ion})=0$. (b) Floquet regime: the probe pulse duration is similar to one optical cycle of the pump field. Note that the probe pulse amplitude is many orders of magnitude weaker than the pump amplitude, but is plotted here on the same scale for clarity.

In terms of extracting physical interpretations from the Tr-ARPES spectra, there could be several contradicting pictures. On the one hand, we have the adiabatic Houston state picture [51,52]. The Houston states are obtained by diagonalizing the instantaneous Hamiltonian, including the pump vector potential only, i.e. within our TDDFT approach, one may write:

$$\left(\frac{1}{2}\left(-i\mathbf{\nabla} + \frac{\mathbf{A}_{\text{pump}}(t')}{c}\right)^2 + v_{KS}(\mathbf{r}, t')\right)|\varphi_{n,\mathbf{k}}^{H-KS}(t')\rangle = \epsilon_{n,\mathbf{k}}^{KS}(t')|\varphi_{n,\mathbf{k}}^{H-KS}(t')\rangle \tag{10}$$

where $|\varphi_{n,\mathbf{k}}^{H-KS}\rangle$ are the adiabatic Houston-KS-Bloch states, with corresponding eigenenergies $\epsilon_{n,\mathbf{k}}^{KS}(t')$, and $t'$ is the instantaneous moment in time where the diagonalization procedure is performed. We emphasize that Eq. (10) is not solved with time propagation of the initial states up to $t'$, but rather, is diagonalized at $t'$ to yield the instantaneous adiabatic band structure (which is time-dependent). If the dynamics are exactly adiabatic, then this procedure should be identical to the time propagation up to an overall phase, the Berry phase [82], which interestingly can be indeed inferred from this picture [83]. The main effects of the Houston picture on the bands are two-fold: (i) the reference crystal momentum $\mathbf{k}$ is translated by the instantaneous pump vector potential, such that $\mathbf{k} \to \mathbf{k}' = \mathbf{k} + \mathbf{A}_{\text{pump}}(t')/c$ [53]. This can be considered equivalent to intra-band acceleration that is a main step in solid-HHG. (ii) The eigenstates can acquire an additional $\mathbf{k}$-dependent, and $t'$-dependent, phase. The acquired phase alters the transition matrix elements from the Bloch states to the continuum, which should modulate the photoemission probability *vs.* $\mathbf{k}$. Most notably, for the moments in time where $\mathbf{A}_{\text{pump}}(t') = 0$, the Houston states correspond exactly to the field-free eigenstates up to the Berry phase, and neglecting effects of dynamical correlations. This suggests an easy procedure to extract deviations from adiabaticity and/or Berry phases, by measuring Tr-ARPES spectra with pump-probe delays for which $\mathbf{A}_{\text{pump}}(t_{ion}) = 0$, and subtracting it from the field-free ARPES spectra. In this context, the degree of adiabaticity of a system is usually estimated with a Keldysh parameter: $\gamma = \omega\sqrt{2E_g}/E_0$, where $E_g$ is the band gap. For $\omega \to 0$ (long-wavelength limit), one might expect that the electrons follow the instantaneous pump field, and the adiabatic picture would hold ($\gamma < 1$). On the other hand, for shorter wavelengths, even if $E_0$ is large, one can expect deviations from adiabaticity ($\gamma > 1$).

On the opposing side to the instantaneous picture, we have the Floquet-picture, which is often employed for analyzing field-driven dynamics [13,14,48,84–87]. Here the effective bands of the field-dressed system correspond to a time-averaged picture. Assuming perfect temporal periodicity in the pump field (neglecting $f(t)$ in Eq. (4)), one can obtain the Floquet-Bloch Hamiltonian [88], which in the TDDFT picture takes the form:

$$H_f^{KS} = \log\left\{\hat{T}\exp\left[-i\frac{\omega}{2\pi}\int_0^{\frac{2\pi}{\omega}}dt\left(\frac{1}{2}\left(-i\mathbf{\nabla} + \frac{\mathbf{A}_{\text{pump}}(t)}{c}\right)^2 + v_{KS}(\mathbf{r}, t)\right)\right]\right\} \tag{11}$$

where $\hat{T}$ is a proper time-ordering operator. The eigenvalues of $H_f^{KS}$ are the Floquet quasi-energies with corresponding Floquet eigenstates (stroboscopic eigenstates). The Floquet states have a time-independent occupation, but evolve dynamically in a time-periodic manner. Due to the cycle-averaging, the quasi-energies are time-independent, which means that in this picture there should not be any pump-probe delay dependence of the Tr-ARPES spectra. Such states can be captured by Tr-ARPES in the Floquet regime outlined above. The implications of the Floquet picture for the strong-field regime of solid state physics remain not fully understood. It has been shown that due to light-dressing, Floquet sideband can be observed in ARPES [89–91] and in the linear optical response [9,10,48,49,92], as well as in charge transport [93,94]. The sidebands are shifted from the original electronic bands by integer multiples of the driving frequency, and are eigenstates of eq. (11). Floquet dressing by phonons is also possible [95]. It has been argued that such dressing could lead to gap closing and topological phases in certain regimes of laser driving [13,14], or even to free-electron like states with a parabolic dispersion [12,55,56]. At the same time, this picture can also be utilized as a basis set for calculating highly nonlinear effects such as HHG. Notably, the Floquet picture is not necessarily at odds with the adiabatic Houston picture – if the dynamics are adiabatic, the Floquet quasi-energy bands could simply emerge as the time-averaged adiabatic Houston bands, where one simply cycle-averages $\epsilon_{n,\mathbf{k}}^{KS}(t')$ in Eq. (10).

In between these pictures, one could obtain hybrid states that are neither Floquet states, nor instantaneous ones. In other words, there could be some diabatic effects in the light-driven electron dynamics. Recent work indicates that such hybrid states leads to gap closings in higher conduction bands [54]. We note that this physical interpretation relies on a different basis expansion of the electronic dynamics, which should give the same results for physical observables as other basis sets. However, since these states have an oscillating form and occupation, they are potentially more difficult to analyze and associate with ARPES spectra.

As a final point, it is also possible that in extremely strong driving, the standard band picture breaks down. This is the expected behavior if one considers the strong-field approximation that is commonly employed for strong light-matter interactions in atoms and molecules [79]. Essentially, one could imagine that $\mathbf{A}_{\text{pump}}(t)$ in Eqs. (10) and (11) might be so strong (because $E_0$ is large, and $\omega$ is small), that it overshadows all other terms in the effective Hamiltonian, leaving the solutions to be approximate eigenstates of the momentum operator (Volkov states). Such behavior would simply lead to a linear dispersion term that is added on top of the field-free bands. Of course, one needs to be careful since that is exactly the same expected dispersion as that of the continuum acceleration of the electrons.

In what follows, we analyze Tr-ARPES from hBN in the strong-field-dressed regime in the context of the above pictures. We will show that, to first order, the instantaneous adiabatic picture prevails as the dominant outlook, even in high field powers and in shorter wavelengths (corresponding to Keldysh parameters up to ~7).

## IV. STRONG-FIELD-DRESSED Tr-ARPES

### A. Instantaneous-regime

We now present the main results of Tr-ARPES in hBN that is pumped by a strong laser field. As a starting point, Fig. 4 presents Tr-ARPES spectra in the instantaneous regime *vs.* pump-probe delay. The pump field power is set at $0.9 \times 10^{12}$ W/cm$^2$ and its wavelength to 2000nm (corresponding to a Keldysh parameter $\gamma \cong 2.5$). These conditions are usually used to generate high harmonics, transient currents, and possibly even novel phases of matter. In our case, the HHG spectrum contains harmonics up to 25eV, validating that both interband and intraband electron dynamics are present during the interaction of hBN with $\mathbf{A}_{\text{pump}}$ (because harmonic emission with energies above $E_g$ can only be generated by excitations to the conduction band and further acceleration in it). Figures 4(a-f) show the ARPES spectra *vs.* pump-probe delay for delays ranging from $\mathbf{A}_{\text{pump}}(t_{ion}) = 0$ (at Fig. 4(a)), up to $\mathbf{A}_{\text{pump}}(t_{ion}) = cE_0/\omega$ (at Fig. 4(f)). Note that we focus on the topmost valence band (that includes the VBM at K and K'), since in these conditions the conduction band occupation is very small and difficult to analyze, and since deeper bands are not expected to be strongly affected by the pump laser because they are strongly bounded.

We highlight several main observations in Fig. 4: (i) The band picture generally holds up in these strong fields. (ii) At all pump-probe delays the bands are very similar to the field-free bands, showing only relatively small deviations. (iii) One main visible effect of the field-dressing is that the valence band (and other deeper bands, not shown) translate along the *k*-axis with the pump-probe delay. This exactly corresponds to the prediction from the adiabatic Houston states – the shift is highlighted in Fig. 4(f) where the pump vector potential is maximized. We verified that this shift indeed corresponds to the value of the vector potential upon the moment of ionization ($\mathbf{A}_{\text{pump}}(t_{ion})$, the moment at which $\mathbf{A}_{\text{probe}}(t)$ is at maxima). (iv) For all delays, one can see a clear asymmetry in the photoemission probability of positive/negative momenta. In other words, the Tr-ARPES spectra are not symmetric under $\mathbf{k} \rightarrow -\mathbf{k}$, even though $\mathbf{A}_{\text{pump}}$ is linearly polarized. For the particular delay in Fig. 4(f), this is a clear result of the Houston state field-dressing that results in a *k*-dependent phase, that in turn modulates the ARPES transition matrix elements. On the other hand, a similar effect is also visible for the delay in Fig. 4(a), for which the instantaneous vector potential is zero at the moment of ionization. Thus, the ARPES transition matrix elements to the continuum provide clear signatures for deviations from the Houston picture, (i.e. for the instantaneous Bloch state field-acquired phases). (v) Lastly, we note that there are a few other effects in the bands that deviate from the Houston state prediction. These will be separately discussed at a later stage.

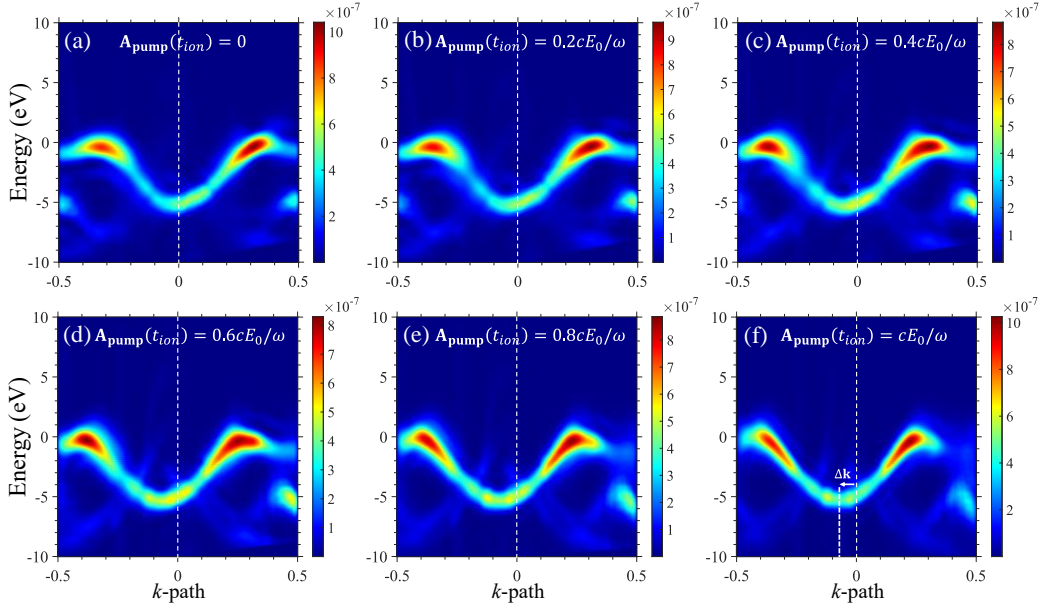

FIG. 4. Tr-ARPES in the instantaneous regime *vs.* pump-probe delay. (a)-(f) Tr-ARPES spectra for different pump-probe delays ranging from **A**$_{\text{pump}}$($t_{ion}$)=0 in (a) to **A**$_{\text{pump}}$($t_{ion}$)= $cE_0/\omega$ in (f) (indicated in the figure). Tr-ARPES spectra are calculated for pump parameters of $I_0$=0.9×10$^{12}$ W/cm$^2$, $\lambda$=2000 nm. The markings in (f) denote a systematic shift of the band center with the pump vector potential, in accordance with the prediction of the adiabatic Houston states (see text). The effects of continuum state dressing have been removed according to the procedure detailed in II.C. Dashed white lines indicate the positions of the Gamma point.

Next, we explore the intensity dependence of the band-dressing effects. Figure 5(a-d) presents the resulting ARPES spectra for a system pumped by 2000nm light, with varying laser powers, at the pump-probe delay $\mathbf{A}_{\text{pump}}(t_{ion}) = 0$. For very weak pump power (Fig. 5(a)), the valence band resembles the field-free system (see Fig. 1(c)). As the pump power is increased, we observe modifications of the bands. Most notably, attenuation of the ARPES transition matrix elements as observed also in Fig. 4. However, we also observe slight bending of the valence band near Γ, the origin of which remains unclear. We hypothesize that this effect arises from non-adiabatic electron dynamics with some short-term memory. One point to keep in mind is that if $\mathbf{A}_{\text{pump}}(t_{ion}) = 0$, then $\mathbf{E}_{\text{pump}}(t_{ion})$ is very large, meaning that strong motion of electrons within the bands could also be the cause for such non-adiabatic effects. Overall, the band shape is similar to the field-free bands in all of these driving conditions, which largely validates the instantaneous Houston band picture. Similar results are obtained for other pump-probe delays.

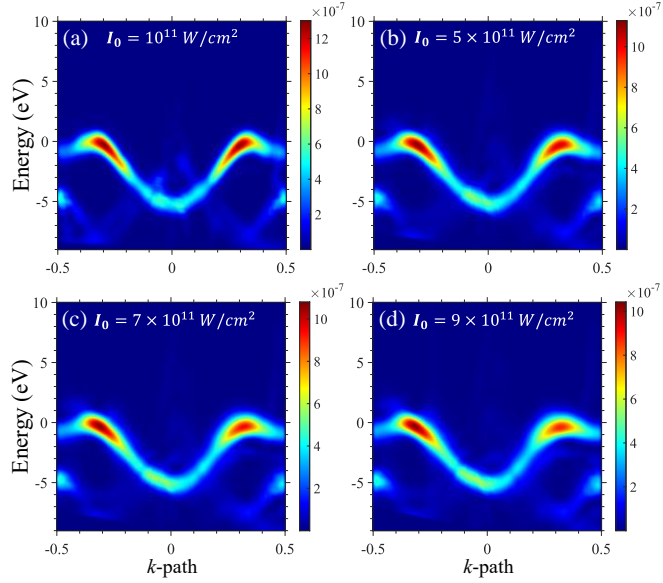

FIG. 5. Tr-ARPES in the instantaneous regime *vs.* pump power for **A***($t_{ion}$)*=0. Pump laser powers are indicated over each figure, and the pump wavelength is taken as *λ*=2000 nm. The effects of continuum state dressing has been removed according to the procedure detailed in II.C.

Figure 6(a-d) explores the ARPES dependence on the pump wavelength (down to 800nm) for strong driving conditions. The probe pulse duration in this case is scaled with the relative pump frequency in order to stay in the instantaneous regime (see appendix for details). This artificially widens the bands due to time-energy Fourier relations, but the main features are still clearly visible. The conclusion arising from Fig. 6 is that the instantaneous adiabatic picture is valid throughout these driving conditions. In fact, we found it largely valid in the domain of Keldysh parameters $\gamma$~1-7, which is well beyond the standardly considered adiabatic limit for the system (with $\gamma < 1$). Of course, Fig. 6 still presents some non-adiabaticity effects that become more pronounced for shorter wavelengths, but the original band shape is preserved. This conclusion is valid for all pump-probe delays, as well as other laser driving conditions studied here.

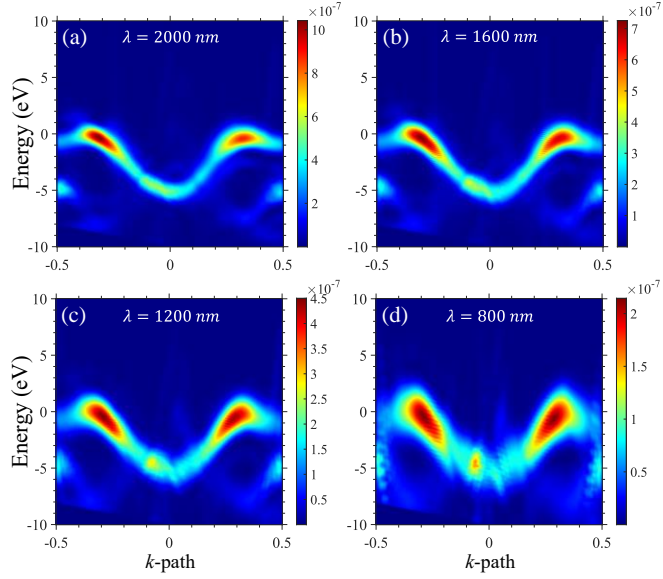

FIG. 6. Tr-ARPES in the instantaneous regime *vs.* pump wavelength for $\mathbf{A}(t_{ion})$=0. The pump wavelengths are indicated in each figure, and the pump power is taken as $I_0$=0.9×10$^{12}$ W/cm$^2$. The effects of continuum state dressing has been removed according to the procedure detailed in II.C.

### B.   Floquet-regime

Up to now, our calculations validate the instantaneous and adiabatic Houston state picture for analyzing field-dressing effects. However, one might object that these results are biased due to the explored parameter regime for $\mathbf{A_{probe}}$. That is, because the XUV pulse probes the instantaneous structure of the system (as it is much shorter than the duration of $\mathbf{A_{pump}}$), a bias towards adiabaticity is created. To address this, we now investigate the Floquet regime, where the probe pulse has a similar duration to one laser cycle of the pump field. In this case, one cannot define a singular moment of ionization for outgoing photoelectrons, because they can be ionized at any moment along the pump laser cycle. As such, all photoemission events are coherently summed in time to obtain the resulting Tr-ARPES spectrum. The time-periodicity of this procedure manifests as Floquet sidebands in the spectra, which arise from destructive/constructive interference of photoemissions from different moments in time. We emphasize that such sidebands are simply a result of the time-periodicity of $\mathbf{A_{pump}}(t)$, and cannot solely validate the Floquet picture for the band-dressing (which can only be done by comparing the measured ARPES spectra to the actual Floquet quasi-energy band structure).

The Floquet regime also leads to some numerical issues: because there is no singular moment of ionization, we cannot fully remove all continuum propagation effects from the spectra (see section II). Even so, we can still determine the origin of the various features appearing in the spectra (band-dressing or continuum) by comparing to Tr-ARPES calculations in the instantaneous regime. We achieve this by performing three levels of calculations: (i) the Tr-ARPES calculations within the Floquet regime described above, which still include partial effects of acceleration in the continuum (within the simulation box). (ii) Tr-ARPES calculations that are calculated in the instantaneous regime *vs.* pump-probe delay, and are averaged incoherently over a full pump laser cycle. This cycle-averaged spectrum is equivalent to the Floquet calculation only as long as the electron dynamics are approximately adiabatic, and up to the phases of the outgoing photoelectrons. Because it lacks the phase information, it will not lead to Floquet sidebands. (iii) A calculation of a cycle-averaged Tr-ARPES in the instantaneous regime just as in (ii), but where the continuum effects are removed. By comparing these different levels of theory, we can conclude what are the main band dressing effects, and as we will show, the adiabatic picture remains valid.

Figure 7 presents calculations of Tr-ARPES spectra with the above-described procedures in the Floquet regime. Figure 7(a) clearly shows the formation of Floquet sidebands with spacing of $\omega$. It also outlines a sizable effect of band widening along the energy-axis, as well as along the *k*-axis. Widening along the *k*-axis can be viewed as the cycle-averaged result of the bands shifting with $\mathbf{A_{pump}}(t)$, as observed also in the instantaneous regime. On the other hand, the band widening in the energy-axis (and consequential gap-closing by ~1 eV) was not observed in the instantaneous regime above, which means that it is either a physical result of Floquet band dressing, or an artifact of continuum propagation. Figure 7(b) shows the cycle-averaged spectrum in the instantaneous regime in similar conditions to Fig. 7(a). Because the spectra are averaged over a cycle incoherently (without phases), no Floquet sidebands appear. Even so, the envelope and main shape of the two spectra are very similar. This indicates that the dynamics are indeed largely adiabatic, since the calculation of Fig. 7(b) explicitly assumes adiabaticity. Note that Fig. 7(b) reproduces the ~1 eV closure of the band gap – a peak in photoemission is found 1eV above the dashed white line that represents the original band edge, whereas standardly such a peak should be centered at the band edge position. Figure 7(c) further presents the cycle-averaged spectrum that is calculated just as in Fig. 7(b), but where continuum effects are subtracted. It directly shows that the gap-closing effect vanishes, verifying that it was an artifact of the continuum acceleration. The band widening along the momentum axis remains, and is a purely adiabatic effect. Similar results are obtained for shorter pump wavelengths. Altogether, we can deduce that the instantaneous adiabatic picture is still widely applicable in these conditions. A cycle-averaged Floquet type of dressing is fully compatible with this Houston state picture, and no novel physical effects are observed.

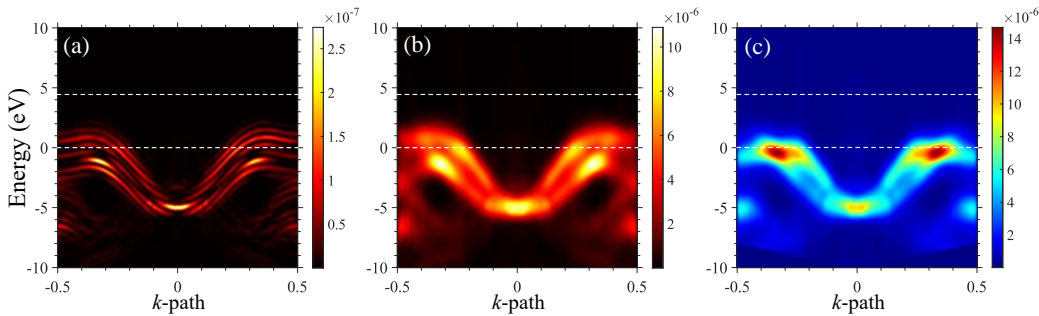

FIG. 7. Tr-ARPES in the Floquet regime with different levels of theory. (a) Tr-ARPES spectrum within the Floquet regime, where the spectrum is not corrected for effects of continuum propagation up to the detector. (b) Tr-ARPES spectrum obtained from incoherently averaging the individual spectra *vs.* delay in the same conditions as in (a), but where the probe pulse is in the instantaneous regime. Here the spectra are averaged over a full laser cycle. Continuum effects of propagation of photoelectron up to the detector are not corrected for. (c) Same as in (b), but where the continuum effects are corrected for. The plots are calculated for pump laser conditions of $I_0=0.9\times10^{12}$ W/cm$^2$ and $\lambda$=2000 nm. Dashed white lines indicate the positions of the valence and conduction band edges.

## V. STRONG-FIELD-DRESSING WITH CO-LINEAR ω-2ω FIELDS

At this point, we focus on the non-adiabatic effects observed above, and outline one possible procedure to probe them more easily. In the following, we replace the monochromatic pump field in Eq. (4) with a two-color co-linearly polarized field:

$$\mathbf{A^{\omega-2\omega}_{pump}}(t) = f(t)\frac{cE_0}{\omega}[\sin(\omega t + \phi) + \Delta\sin(2\omega t)]\mathbf{\hat{e}} \qquad (12)$$

where $\phi$ is the relative phase between the two beams, and $\Delta$ is the relative amplitude between the beams. $\mathbf{A^{\omega-2\omega}_{pump}}(t)$ describes a two-color coherent field that effectively breaks spatial inversion and time-reversal symmetries [87]. Let us explain why this symmetry-breaking is important in the context of non-adiabatic band effects. Non-adiabaticity is most easily observed in pump-probe delays for which $\mathbf{A_{pump}}(t_{ion}) = 0$, because for those delays the adiabatic picture predicts no modifications (assuming that the contributions of dynamical correlations is negligible). When using a monochromatic field as in Eq. (4), the vector potential vanishes twice per optical cycle (i.e. for every cycle of the pump field there are two pump-probe delays that yield $\mathbf{A_{pump}}(t_{ion}) = 0$). Both of these moments are connected by symmetry. In one of them $E_{pump}(t_{ion}) > 0$, while in the other $E_{pump}(t_{ion}) < 0$, but the absolute value of the field is identical in both. Consequently, the ARPES spectra are identical up to a mirror image – the time-reversal and mirror symmetries in $\mathbf{A_{pump}}$ connect K to K'. On the other hand, if those spectra were different, one could compare the deviations between them and search for diabatic effects in the dressing dynamics. This is exactly what the two-color field in Eq. (12) and its symmetry breaking implies – there can be up to four temporal moments in a single laser cycle for which $\mathbf{A_{pump}^{\omega-2\omega}}(t_{ion}) = 0$. Moreover, even though $\mathbf{A_{pump}^{\omega-2\omega}} = 0$ in each of those moments, the instantaneous electric field values, $\mathbf{E_{pump}^{\omega-2\omega}}(t_{ion})$, can be widely different, disconnecting K and K'. This provides additional freedom for observing deviations from the adiabatic bands, especially while scanning the relative beam phase and amplitude.

Figures 8 and 9 present this analysis for two cases. In Fig. 8 we employ Δ=1, and $\phi=\pi/4$. This leads to a vector potential that has two zeros per optical cycle, same as in the monochromatic case (see Fig. 8(a)), but with different values for the instantaneous electric fields. Another noteworthy point is that the 'history' of the vector potential's behavior (and potentially the corresponding electron dynamics) before those particular moments in time is different between these two temporal moments, since the field itself has broken time-reversal symmetry. Indeed from Figs. 8(b,c) we observe that the two spectra are not mirror images. Deviations between the images are clear at the K and K' points, where the band curvatures slightly differ, and where the ARPES amplitude differs. These are highlighted in Fig. 8(d), which plots the trace of the ARPES spectra for both time-delays at a fixed energy, corresponding to the valence band edge. Note that in Fig. 8(d) the plot is reflected with $\mathbf{k} \rightarrow -\mathbf{k}$ for one of the delays, which allows directly comparing deviations from mirror symmetry. Such effects could be fingerprints of short-term memory field-dressing electron dynamics, which are easier to observe with the two-color analysis. We note that here $E_0$ in Eq. (8) is scaled down to correspond to a power of $2.5 \times 10^{11}$ W/cm$^2$, such that the total maximal field amplitude (from the coherent superposition of both beams) is close to that used in the previous sections. Moreover, the wavelength of the fundamental period is taken as 4000nm to make sure that any non-adiabaticity is on equal footing to previous sections (the fastest frequency component in the beam is at 2000nm).

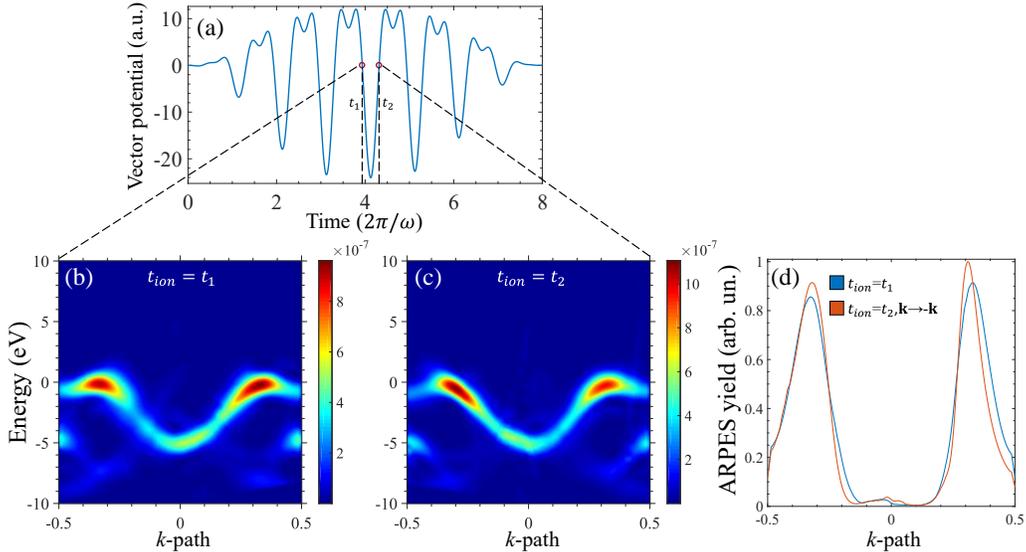

FIG. 8. Tr-ARPES in the instantaneous regime with an ω-2ω bi-chromatic pump (eq. (12)) for Δ=1 and $\phi=\pi/4$. (a) Total vector potential *vs.* time showing that the laser field induces time-reversal and inversion symmetry breaking. Two moments in time where the vector potential is zero are highlighted with dashed lines and denoted by $t_1$ and $t_2$. While the vector potential is instantaneously zero in both of these moments, the evolution of the field prior to them is different, and so is the instantaneous electric field. (b) Tr-ARPES spectrum corresponding to a pump-probe delay where the peak ionization occurs at $t_1$. (c) same as (b), but for $t_2$. Plots calculated with pump parameters λ=4000 nm and $I_0=2.5\times10^{11}$ W/cm$^2$. The effects of continuum state dressing has been removed according to the procedure detailed in II.C. (d) ARPES yield traces for a fixed photoelectron energy at the VBM (along the dashed white lines in (b) and (c)), for the two pump-probe delays, but where for $t_2$ the plot is reflected with $\mathbf{k} \rightarrow -\mathbf{k}$. For a monochromatic field, these two curves are identical, whereas the deviation indicates nonadiabatic effects accessible with the two-color approach.

Figure 9 addresses a regime with $\Delta = \sqrt{2}$, and $\phi = \pi/2$ (the total power is similarly scaled to $2\times10^{11}$ W/cm$^2$ and the fundamental wavelength to 4000nm to avoid any bias towards enhanced non-adiabaticity). For these parameters there are four zeros in $\mathbf{A}_{\text{pump}}^{\omega-2\omega}$ in each laser cycle, which correspond to completely different behaviors in the pumping electric field (see Fig. 9(a) with the highlighted moments in time). The resulting Tr-ARPES spectra for the corresponding delays are widely different: Figures 9(c) and (d) are remarkably similar to the field-free bands, while Figs. 9(b) and (e) show stronger modifications. Especially, Fig. 9(e) shows strong modifications of the band curvature near K and K'. Since Fig. 9(e) corresponds to the largest instantaneous electric field of these delays, and also shows the largest modifications, it suggests that an extension of the adiabatic theory to include the instantaneous electric field (the temporal derivative of $\mathbf{A}_{\text{pump}}(t)$ that also incorporates some non-temporally-local effects) might improve the interpretation. Notably, in the Appendix we show that in these conditions dynamical correlations and *e-e* interactions contribute negligibly to the ARPES spectra, and are thus not a main source for the observed deviations from adiabaticity. Such investigations should be topics of future work.

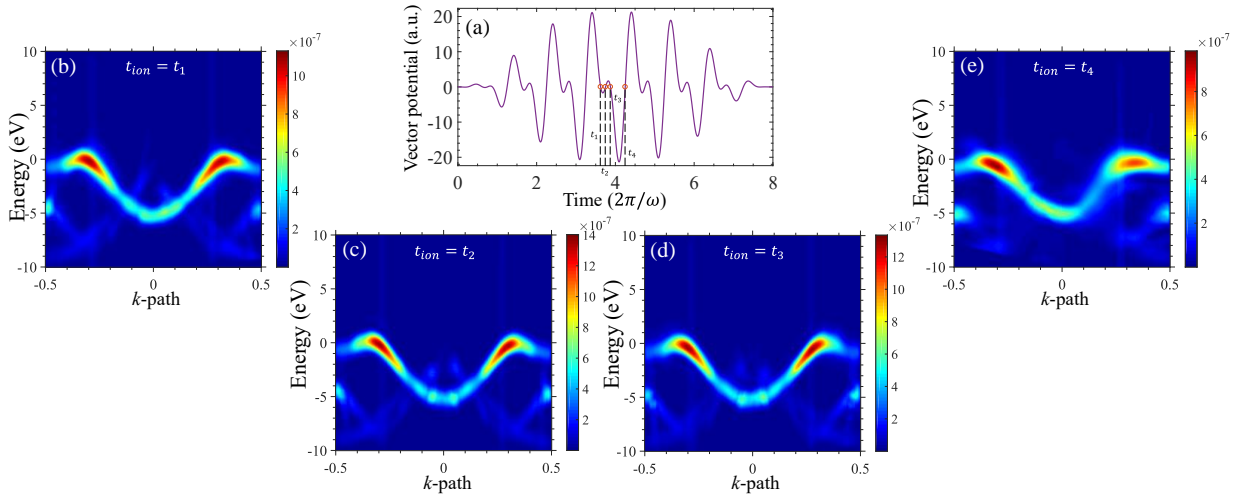

FIG. 9. Tr-ARPES in the instantaneous regime with an ω-2ω bi-chromatic pump (eq. (12)) for $\Delta=\sqrt{2}$ and $\phi=\pi/2$. (a) Total vector potential *vs.* time showing that the laser field induces time-reversal and inversion symmetry breaking. For these parameters there are four moments in time where the vector potential is zero, which are highlighted with dashed lines and denoted by $t_{1-4}$. While the vector potential is instantaneously zero all of these moments, the evolution of the field prior to them is different, and so is the instantaneous electric field. (b)-(e) Tr-ARPES spectra corresponding to a pump-probe delay where the peak ionization occurs at $t_1$, $t_2$, $t_3$, and $t_4$, respectively. Plots calculated with pump parameters λ=4000 nm and $I_0=2\times10^{11}$ W/cm$^2$. The effects of continuum state dressing has been removed according to the procedure detailed in II.C. Note that there is some numerical noise around K and K' points for these strong driving conditions causing the vertical lines in the ARPES spectra.

## VI. SUMMARY

To summarize, we explored here Tr-ARPES in a benchmark solid system that is irradiated by a strong pump laser. The resulting ARPES spectra allow to directly, and without additional assumptions, observe field-dressed states of matter. This opens the path towards resolving unanswered questions in strong-field physics of solids. For hBN, we determined that the adiabatic bands, as predicted by the Houston state picture [51,52], are well valid for a very wide parameter regime. In particular, only weak modifications of the bands were observed in laser powers up to $10^{12}$ W/cm$^2$, and down to laser wavelength of 800nm. Surprisingly, the corresponding Keldysh parameter ranges are up to 7, which is much above the standard notion of adiabaticity in light-matter systems. We further validated that this conclusion is upheld even in the Floquet regime, where the ARPES spectra are comprised from coherent motion of electrons in the bands over a full pump laser cycle. Lastly, we put forw0061rd a time-resolved approach for observing non-adiabatic effects in bands by measuring Tr-ARPES spectra at pump-probe delays where the instantaneous pump vector potential is zero. With this technique, we showed that there are some interesting effects of band modifications due to light-dressing, which can be sensitively probed by pumping the system with a bi-chromatic ω-2ω field. These results suggest that the adiabatic theory could potentially be amended by adding a term that correlates with the instantaneous electric field, on top of the instantaneous vector potential.

Our work further validates that the adiabatic band picture should be the correct approach for interpreting strong-field driven dynamics in solids, at least in most cases where the pumping photon energy is far from resonances. This result provides theoretical soundness behind recent efforts for band structure and Berry curvature reconstructions in various systems [43–45,47]. At the same time, it remains unclear under what conditions strong modifications, or breaking, of solid bands could be observed in this regime, as we have not found any indication for it in the our simulations. Looking forward, several extensions of this approach present interesting prospects. First, it would be interesting to apply this technique to explore topological systems [20–26], systems driven by topological and chiral light [96–101], strongly-correlated systems [28,30,31], nonlinear photocurrents [3–8], systems experiencing symmetry breaking [102,103], and more. Second, Tr-ARPES in the strong-field regime, and instantaneous picture, presents a unique opportunity to measure photoemission time-delays [80,81] to high accuracy – because one can technically time the moment of ionization by mapping it to the instantaneous shifting of the band structure with the pump vector potential. Further development along these lines should be topic of future work.

## ACKNOWLEDGEMENTS

We acknowledge financial support from the European Research Council (ERC-2015-AdG-694097). This work was supported by the Cluster of Excellence Advanced Imaging of Matter (AIM), Grupos Consolidados (IT1249-19) and SFB925. The Flatiron Institute is a division of the Simons Foundation. O.N. gratefully acknowledges the generous support of a Schmidt Science Fellowship.

## APPENDIX

We provide here additional technical details for the calculations performed in the main text, as well as some complementary results. The ground state DFT calculations were performed with octopus code [69–72]. The KS states were discretized on a Cartesian grid with a hexagonal unit cell in the *xy* plane, corresponding to the primitive cell of monolayer hBN. The transverse *z*-axis was taken to have a total length of 140 Bohr, with 70 Bohr vacuum spacing on each side of the monolayer. Lattice parameters and atomic positions were taken at experimental values, and grid spacing was taken to be 0.4 Bohr in all directions. We employed semi-periodic boundary conditions with two periodic dimensions in the monolayer plane (along the lattice vectors), and a non-periodic *z*-axis, with a 12x12x1 Γ-centered *k*-grid. Spin degrees of freedom (as well as spin-orbit coupling) were neglected. Deep core bands were replaced with norm-conserving pseudopotentials [104]. The KS equations were solved self-consistently with a tolerance of <$10^{-7}$ Hartree. Results were converged with respect to grid spacing and *k*-grid density.

For TDDFT calculations, we employed a complex absorbing boundary in addition to the KS potential described in eq. (2), with a width of 40 Bohr along the *z*-axis, on both sides of the monolayer, and a $\sin^2(z)$ shape that saturates to a height of -1. This effectively means that photoelectrons have a distance of 30 Bohr to traverse before reaching the onset of the complex absorbing potential, where the numerical detector is also located ($w$=30 Bohr), and the photoelectron momentum-resolved flux is computed for obtaining ARPES spectra. We employed a time step of $\Delta t$ =0.07 a.u. for the propagation scheme.

The ARPES spectra were obtained with T-SURFF method as implemented in Octopus [74,75], but where the propagation of the outgoing waves was not performed beyond the numerical detector at $z = w$. This is done in order to partially remove effects of propagation in the continuum (see main text). The only exception is Fig. 2(a), where a 'standard' T-SURFF approach was utilized including propagation of the Volkov states with the full vector potential. The ARPES spectra was calculated over an energy grid with spacing of 0.01 eV, and over a separate *k*-grid with 72 *k*-points traversing from (-0.5,-0.5,0) up to (0.5,0.5,0) in reciprocal space (in fractional units of the reciprocal lattice vectors). The resulting spectra were then smoothed with a moving mean filter.

The envelope function of the employed pump laser pulse, *f(t)*, was taken to be of the following 'super-sine' form [105]:

$$f(t) = \left(sin\left(\pi \frac{t}{T_p}\right)\right)^{\left(\frac{\left|\pi\left(\frac{t}{T_p}-\frac{1}{2}\right)\right|}{\sigma}\right)} \tag{13}$$

where $\sigma=0.75$, $T_p$ is the duration of the laser pulse which was taken to be $T_p=8T$, where $T$ is a single cycle of the fundamental carrier frequency ($T = 2\pi/\omega$). This form is roughly analogous to a super-gaussian pulse, but where the $f(t)$ starts and ends exactly at 0 which is more convenient numerically. The corresponding full-width-half-max (FWHM) of the pulse is $4T$. The envelope function for the probe laser pulse was taken to have a similar form:

$$f_{xuv}(t) = \left(sin\left(\pi \frac{t}{T_{xuv}}\right)\right)^{\left(\frac{\left|\pi\left(\frac{t}{T_{xuv}}-\frac{1}{2}\right)\right|}{\sigma}\right)} \tag{14}$$

where $T_{xuv}$ is the total duration of the probe pulse. For the instantaneous regime calculations we employed $T_{xuv} = 100(2\pi/\omega_{xuv})$, which has a FWHM of 2.1 femtoseconds. For the calculations in Fig. 6 this duration was linearly scaled with the pump wavelength, i.e., $T_{xuv} \rightarrow T_{xuv}' = \left(\frac{\lambda}{2000nm}\right)T_{xuv}$. For calculations in the Floquet regime performed in Fig. 7(a), we employed $T_{xuv} = 400(2\pi/\omega_{xuv})$, with a FWHM of 8.2 femtoseconds (which is a little longer than the pump period at 2000nm which is 6.7 femtoseconds.

Lastly, we show here that the results obtained in the main text are largely independent of dynamical correlations and *e-e* interactions by comparing results obtained from the full TDDFT calculations (as presented in the main text), to calculations within the independent particle approximation (IPA). In the IPA we freeze the dynamical evolution of the Hartree and XC terms in the Hamiltonian, which decouples the equations of motion of all the KS-Bloch states, and removes any dynamical correlations from the simulations. Such effects could also be considered as a source for diabatic effects, regardless of light-dressed states. Figure 10 presents an exemplary calculation of an ARPES spectra in the instantaneous regime with the IPA compared to the full TDDFT calculation. The two calculations match extremely well, and similar results are obtained for other laser powers. Thus, we conclude that dynamical correlations are likely not a source of the main deviations from adiabaticity in the bands and that dynamical *e-e* interactions do not substantially effect the Tr-ARPES spectra.

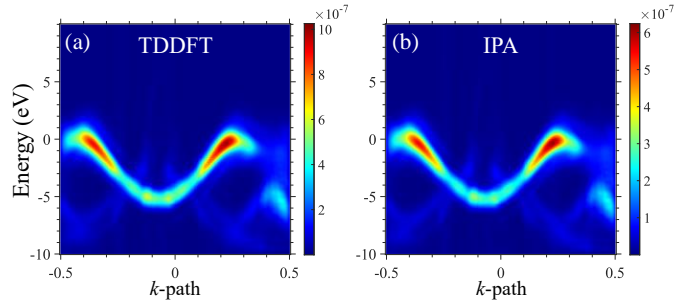

FIG. 10. Tr-ARPES in the instantaneous regime comparing the IPA to a full TDDFT calculation. In both ARPES spectra the system is driven by a pump field with a wavelength of 2000nm and $I_0=9\times10^{11}$ W/cm$^2$, and the pump-probe delay is set such that $\mathbf{A_{pump}}(t_{ion})= cE_0/\omega$. (a) Full TDDFT calculation. (b) IPA calculation.

## REFERENCES

[1]  S. Ghimire, A. D. Dichiara, E. Sistrunk, P. Agostini, L. F. DiMauro, and D. A. Reis, *Observation of High-Order Harmonic Generation in a Bulk Crystal*, Nat. Phys. **7**, 138 (2011).

[2]  S. Ghimire and D. A. Reis, *High-Harmonic Generation from Solids*, Nat. Phys. **15**, 10 (2019).

[3]  A. Schiffrin, T. Paasch-Colberg, N. Karpowicz, V. Apalkov, D. Gerster, S. Mühlbrandt, M. Korbman, J. Reichert, M. Schultze, S. Holzner, J. V Barth, R. Kienberger, R. Ernstorfer, V. S. Yakovlev, M. I. Stockman, and F. Krausz, *Optical-Field-Induced Current in Dielectrics*, Nature **493**, 70 (2013).

[4]  T. Higuchi, C. Heide, K. Ullmann, H. B. Weber, and P. Hommelhoff, *Light-Field-Driven Currents in Graphene*, Nature **550**, 224 (2017).

[5]  C. Heide, T. Higuchi, H. B. Weber, and P. Hommelhoff, *Coherent Electron Trajectory Control in Graphene*, Phys. Rev. Lett. **121**, 207401 (2018).

[6]  F. Langer, Y.-P. Liu, Z. Ren, V. Flodgren, C. Guo, J. Vogelsang, S. Mikaelsson, I. Sytcevich, J. Ahrens, A. L'Huillier, C. L. Arnold, and A. Mikkelsen, *Few-Cycle Lightwave-Driven Currents in a Semiconductor at High Repetition Rate*, Optica **7**, 276 (2020).

[7]  S. Sederberg, F. Kong, F. Hufnagel, C. Zhang, E. Karimi, and P. B. Corkum, *Vectorized Optoelectronic Control and Metrology in a Semiconductor*, Nat. Photonics **14**, 680 (2020).

[8]  O. Neufeld, N. Tancogne-Dejean, U. De Giovannini, H. Hübener, and A. Rubio, *Light-Driven Extremely Nonlinear Bulk Photogalvanic Currents*, Phys. Rev. Lett. **127**, 126601 (2021).

[9]  M. Lucchini, S. A. Sato, A. Ludwig, J. Herrmann, M. Volkov, L. Kasmi, Y. Shinohara, K. Yabana, L. Gallmann, and U. Keller, *Attosecond Dynamical Franz-Keldysh Effect in Polycrystalline Diamond*, Science **353**, 916 (2016).

[10] T. Otobe, *Analytical Formulation for Modulation of Time-Resolved Dynamical Franz-Keldysh Effect by Electron Excitation in Dielectrics*, Phys. Rev. B **96**, 235115 (2017).

[11] Q. T. Vu, H. Haug, O. D. Mücke, T. Tritschler, M. Wegener, G. Khitrova, and H. M. Gibbs, *Light-Induced Gaps in Semiconductor Band-to-Band Transitions*, Phys. Rev. Lett. **92**, 217403 (2004).

[12] H. Lakhotia, H. Y. Kim, M. Zhan, S. Hu, S. Meng, and E. Goulielmakis, *Laser Picoscopy of Valence Electrons in Solids*, Nature **583**, 55 (2020).

[13] Á. Jiménez-Galán, R. E. F. Silva, O. Smirnova, and M. Ivanov, *Lightwave Control of Topological Properties in 2D Materials for Sub-Cycle and Non-Resonant Valley Manipulation*, Nat. Photonics **14**, 728 (2020).

[14] N. H. Lindner, G. Refael, and V. Galitski, *Floquet Topological Insulator in Semiconductor Quantum Wells*, Nat. Phys. **7**, 490 (2010).

[15] D. Shin, H. Hübener, U. De Giovannini, H. Jin, A. Rubio, and N. Park, *Phonon-Driven Spin-Floquet Magneto-Valleytronics in MoS2*, Nat. Commun. **9**, 638 (2018).

[16] F. Langer, C. P. Schmid, S. Schlauderer, M. Gmitra, J. Fabian, P. Nagler, C. Schüller, T. Korn, P. G. Hawkins, J. T. Steiner, U. Huttner, S. W. Koch, M. Kira, and R. Huber, *Lightwave Valleytronics in a Monolayer of Tungsten Diselenide*, Nature **557**, 76 (2018).

[17] Á. Jiménez-Galán, R. E. F. Silva, O. Smirnova, and M. Ivanov, *Sub-Cycle Valleytronics: Control of Valley Polarization Using Few-Cycle Linearly Polarized Pulses*, Optica **8**, 277 (2021).

[18] M. S. Mrudul, Á. Jiménez-Galán, M. Ivanov, and G. Dixit, *Light-Induced Valleytronics in Pristine Graphene*, Optica **8**, 422 (2021).

[19] J. Schoetz, Z. Wang, E. Pisanty, M. Lewenstein, M. F. Kling, and M. F. Ciappina, *Perspective on Petahertz Electronics and Attosecond Nanoscopy*, ACS Photonics **6**, 3057 (2019).

[20] D. Bauer and K. K. Hansen, *High-Harmonic Generation in Solids with and without Topological Edge States*, Phys. Rev. Lett. **120**, 177401 (2018).

[21] R. E. F. Silva, Á. Jiménez-Galán, B. Amorim, O. Smirnova, and M. Ivanov, *Topological Strong-Field Physics on Sub-Laser-Cycle Timescale*, Nat. Photonics **13**, 849 (2019).

[22] A. Chacón, D. Kim, W. Zhu, S. P. Kelly, A. Dauphin, E. Pisanty, A. S. Maxwell, A. Picón, M. F. Ciappina, D. E. Kim, C. Ticknor, A. Saxena, and M. Lewenstein, *Circular Dichroism in Higher-Order Harmonic Generation: Heralding Topological Phases and Transitions in Chern Insulators*, Phys. Rev. B **102**, 134115 (2020).

[23] Y. Bai, F. Fei, S. Wang, N. Li, X. Li, F. Song, R. Li, Z. Xu, and P. Liu, *High-Harmonic Generation from Topological Surface States*, Nat. Phys. **17**, 311 (2021).

[24] D. Baykusheva, A. Chacón, D. Kim, D. E. Kim, D. A. Reis, and S. Ghimire, *Strong-Field Physics in Three-Dimensional Topological Insulators*, Phys. Rev. A **103**, 23101 (2021).

[25] C. P. Schmid, L. Weigl, P. Grössing, V. Junk, C. Gorini, S. Schlauderer, S. Ito, M. Meierhofer, N. Hofmann, D. Afanasiev, J. Crewse, K. A. Kokh, O. E. Tereshchenko, J. Güdde, F. Evers, J. Wilhelm, K. Richter, U. Höfer, and R. Huber, *Tunable Non-Integer High-Harmonic Generation in a Topological Insulator*, Nature **593**, 385 (2021).

[26] D. Baykusheva, A. Chacón, J. Lu, T. P. Bailey, J. A. Sobota, H. Soifer, P. S. Kirchmann, C. Rotundu, C. Uher, T. F. Heinz, D. A. Reis, and S. Ghimire, *All-Optical Probe of Three-Dimensional Topological Insulators Based on High-Harmonic Generation by Circularly Polarized Laser Fields*, Nano Lett. **21**, 8970 (2021).

[27] Niccolò Baldelli, U. Bhattacharya, D. González-Cuadra, M. Lewenstein, and T. Graß, *Detecting Majorana Zero Modes via Strong Field Dynamics*, ArXiv:2202.03547 (2022).

[28] R. E. F. Silva, I. V Blinov, A. N. Rubtsov, O. Smirnova, and M. Ivanov, *High-Harmonic Spectroscopy of Ultrafast Many-Body Dynamics in Strongly Correlated Systems*, Nat. Photonics **12**, 266 (2018).

[29] Y. Murakami, M. Eckstein, and P. Werner, *High-Harmonic Generation in Mott Insulators*, Phys. Rev. Lett. **121**, 57405 (2018).

[30] N. Tancogne-Dejean, M. A. Sentef, and A. Rubio, *Ultrafast Modification of Hubbard U in a Strongly Correlated Material: Ab Initio High-Harmonic Generation in NiO*, Phys. Rev. Lett. **121**, 097402 (2018).

[31] G. McCaul, C. Orthodoxou, K. Jacobs, G. H. Booth, and D. I. Bondar, *Driven Imposters: Controlling Expectations in Many-Body Systems*, Phys. Rev. Lett. **124**, 183201 (2020).

[32] C. Shao, H. Lu, X. Zhang, C. Yu, T. Tohyama, and R. Lu, *High-Harmonic Generation Approaching the Quantum Critical Point of Strongly Correlated Systems*, Phys. Rev. Lett. **128**, 47401 (2022).

[33] G. Vampa, C. R. McDonald, G. Orlando, D. D. Klug, P. B. Corkum, and T. Brabec, *Theoretical Analysis of High-Harmonic Generation in Solids*, Phys. Rev. Lett. **113**, 073901 (2014).

[34] O. Schubert, M. Hohenleutner, F. Langer, B. Urbanek, C. Lange, U. Huttner, D. Golde, T. Meier, M. Kira, S. W. Koch, and R. Huber, *Sub-Cycle Control of Terahertz High-Harmonic Generation by Dynamical Bloch Oscillations*, Nat. Photonics **8**, 119 (2014).

[35] M. Wu, S. Ghimire, D. A. Reis, K. J. Schafer, and M. B. Gaarde, *High-Harmonic Generation from Bloch Electrons in Solids*, Phys. Rev. A - At. Mol. Opt. Phys. **91**, (2015).

[36] G. Vampa, T. J. Hammond, N. Thire, B. E. Schmidt, F. Legare, C. R. McDonald, T. Brabec, and P. B. Corkum, *Linking High Harmonics from Gases and Solids*, Nature **522**, (2015).

[37] M. Wu, D. A. Browne, K. J. Schafer, and M. B. Gaarde, *Multilevel Perspective on High-Order Harmonic Generation in Solids*, Phys. Rev. A **94**, 1 (2016).

[38] T. Ikemachi, Y. Shinohara, T. Sato, J. Yumoto, M. Kuwata-Gonokami, and K. L. Ishikawa, *Trajectory Analysis of High-Order-Harmonic Generation from Periodic Crystals*, Phys. Rev. A **95**, 1 (2017).

[39] E. N. Osika, A. Chacón, L. Ortmann, N. Suárez, J. A. Pérez-Hernández, B. Szafran, M. F. Ciappina, F. Sols, A. S. Landsman, and M. Lewenstein, *Wannier-Bloch Approach to Localization in High-Harmonics Generation in Solids*, Phys. Rev. X **7**, 1 (2017).

[40] C. Yu, S. Jiang, and R. Lu, *High Order Harmonic Generation in Solids: A Review on Recent Numerical Methods*, Adv. Phys. X **4**, 1562982 (2019).

[41] L. Li, P. Lan, X. Zhu, and P. Lu, *Huygens-Fresnel Picture for High Harmonic Generation in Solids*, Phys. Rev. Lett. **127**, 223201 (2021).

[42] L. Yue and M. B. Gaarde, *Introduction to Theory of High-Harmonic Generation in Solids: Tutorial*, J. Opt. Soc. Am. B **39**, 535 (2022).

[43] G. Vampa, T. J. Hammond, N. Thiré, B. E. Schmidt, F. Légaré, C. R. McDonald, T. Brabec, D. D. Klug, and P. B. Corkum, *All-Optical Reconstruction of Crystal Band Structure*, Phys. Rev. Lett. **115**, 193603 (2015).

[44] A. A. Lanin, E. A. Stepanov, A. B. Fedotov, and A. M. Zheltikov, *Mapping the Electron Band Structure by Intraband High-Harmonic Generation in Solids*, Optica **4**, 516 (2017).

[45] T. T. Luu and H. J. Wörner, *Measurement of the Berry Curvature of Solids Using High-Harmonic Spectroscopy*, Nat. Commun. **9**, 916 (2018).

[46] A. J. Uzan, G. Orenstein, Á. Jiménez-Galán, C. McDonald, R. E. F. Silva, B. D. Bruner, N. D. Klimkin, V. Blanchet, T. Arusi-Parpar, M. Krüger, A. N. Rubtsov, O. Smirnova, M. Ivanov, B. Yan, T. Brabec, and N. Dudovich, *Attosecond Spectral Singularities in Solid-State High-Harmonic Generation*, Nat. Photonics **14**, 183 (2020).

[47] Y.-Y. Lv, J. Xu, S. Han, C. Zhang, Y. Han, J. Zhou, S.-H. Yao, X.-P. Liu, M.-H. Lu, H. Weng, Z. Xie, Y. B. Chen, J. Hu, Y.-F. Chen, and S. Zhu, *High-Harmonic Generation in Weyl Semimetal β-WP2 Crystals*, Nat. Commun. **12**, 6437 (2021).

[48] M. Holthaus, *Floquet Engineering with Quasienergy Bands of Periodically Driven Optical Lattices*, J. Phys. B At. Mol. Opt. Phys. **49**, 13001 (2016).

[49] T. Otobe, Y. Shinohara, S. A. Sato, and K. Yabana, *Femtosecond Time-Resolved Dynamical Franz-Keldysh Effect*, Phys. Rev. B **93**, 45124 (2016).

[50] S. A. Sato, H. Hübener, and U. De Giovannini, *Applied Sciences Ab Initio Simulation of Attosecond Transient Absorption Spectroscopy in Two-Dimensional Materials*, (2018).

[51] W. V Houston, *Acceleration of Electrons in a Crystal Lattice*, Phys. Rev. **57**, 184 (1940).

[52] J. B. Krieger and G. J. Iafrate, *Time Evolution of Bloch Electrons in a Homogeneous Electric Field*, Phys. Rev. B **33**, 5494 (1986).

[53] V. Gruzdev and O. Sergaeva, *Ultrafast Modification of Band Structure of Wide-Band-Gap Solids by Ultrashort Pulses of Laser-Driven Electron Oscillations*, Phys. Rev. B **98**, 115202 (2018).

[54] A. Uzan-Narovlansky, A. Jimenez-Galan, G. Orenstein, R. Silva, T. Arusi-Parpar, S. Shames, B. Bruner, B. Yan, O. Smirnova, M. (Mikhail) Ivanov, and N. Dudovich, *Observation of Light Induced Dynamical Band Structure via Multi-Band High Harmonic Spectroscopy*, Nat. Portf. (2022).

[55] N. Tzoar and J. I. Gersten, *Theory of Electronic Band Structure in Intense Laser Fields*, Phys. Rev. B **12**, 1132 (1975).

[56] L. C. M. Miranda, *Energy-Gap Distortion in Solids under Intense Laser Fields*, Solid State Commun. **45**, 783 (1983).

[57] G. L. Dakovski, Y. Li, T. Durakiewicz, and G. Rodriguez, *Tunable Ultrafast Extreme Ultraviolet Source for Time- and Angle-Resolved Photoemission Spectroscopy*, Rev. Sci. Instrum. **81**, 73108 (2010).

[58] S. Eich, A. Stange, A. V Carr, J. Urbancic, T. Popmintchev, M. Wiesenmayer, K. Jansen, A. Ruffing, S. Jakobs, T. Rohwer, S. Hellmann, C. Chen, P. Matyba, L. Kipp, K. Rossnagel, M. Bauer, M. M. Murnane, H. C. Kapteyn, S. Mathias, and M. Aeschlimann, *Time- and Angle-Resolved Photoemission Spectroscopy with Optimized High-Harmonic Pulses Using Frequency-Doubled Ti:Sapphire Lasers*, J. Electron Spectros. Relat. Phenomena **195**, 231 (2014).

[59] M. Puppin, Y. Deng, C. W. Nicholson, J. Feldl, N. B. M. Schröter, H. Vita, P. S. Kirchmann, C. Monney, L. Rettig, M. Wolf, and R. Ernstorfer, *Time- and Angle-Resolved Photoemission Spectroscopy of Solids in the Extreme Ultraviolet at 500 KHz Repetition Rate*, Rev. Sci. Instrum. **90**, 23104 (2019).

[60] A. Damascelli, *Probing the Electronic Structure of Complex Systems by ARPES*, Phys. Scr. **T109**, 61 (2004).

[61] B. Lv, T. Qian, and H. Ding, *Angle-Resolved Photoemission Spectroscopy and Its Application to Topological Materials*, Nat. Rev. Phys. **1**, 609 (2019).

[62] R. A., S. R. Majlin, M. M., R. P. K., F. M. H., R. J., C. K., D. Y., and K. A., *Chiral Superconductivity in the Alternate Stacking Compound 4Hb-TaS2*, Sci. Adv. **6**, eaax9480 (2022).

[63] S. Beaulieu, J. Schusser, S. Dong, M. Schüler, T. Pincelli, M. Dendzik, J. Maklar, A. Neef, H. Ebert, K. Hricovini, M. Wolf, J. Braun, L. Rettig, J. Minár, and R. Ernstorfer, *Revealing Hidden Orbital Pseudospin Texture with Time-Reversal Dichroism in Photoelectron Angular Distributions*, Phys. Rev. Lett. **125**, 216404 (2020).

[64] S. Beaulieu, M. Schüler, J. Schusser, S. Dong, T. Pincelli, J. Maklar, A. Neef, F. Reinert, M. Wolf, L. Rettig, J. Minár, and R. Ernstorfer, *Unveiling the Orbital Texture of 1T-TiTe2 Using Intrinsic Linear Dichroism in Multidimensional Photoemission Spectroscopy*, Npj Quantum Mater. **6**, 93 (2021).

[65] H. Soifer, A. Gauthier, A. F. Kemper, C. R. Rotundu, S.-L. Yang, H. Xiong, D. Lu, M. Hashimoto, P. S. Kirchmann, J. A. Sobota, and Z.-X. Shen, *Band-Resolved Imaging of Photocurrent in a Topological Insulator*, Phys. Rev. Lett. **122**, 167401 (2019).

[66] U. De Giovannini, H. Hübener, S. A. Sato, and A. Rubio, *Direct Measurement of Electron-Phonon Coupling with Time-Resolved ARPES*, Phys. Rev. Lett. **125**, 136401 (2020).

[67] P. Hein, S. Jauernik, H. Erk, L. Yang, Y. Qi, Y. Sun, C. Felser, and M. Bauer, *Mode-Resolved Reciprocal Space Mapping of Electron-Phonon Interaction in the Weyl Semimetal Candidate Td-WTe2*, Nat. Commun. **11**, 2613 (2020).

[68] T. Suzuki, Y. Shinohara, Y. Lu, M. Watanabe, J. Xu, K. L. Ishikawa, H. Takagi, M. Nohara, N. Katayama, H. Sawa, M. Fujisawa, T. Kanai, T. Itatani, T. Mizokawa, S. Shin, and K. Okazaki, *Detecting Electron-Phonon Coupling during Photoinduced Phase Transition*, Phys. Rev. B **103**, L121105 (2021).

[69] M. A. L. Marques, A. Castro, G. F. Bertsch, and A. Rubio, *Octopus: A First-Principles Tool for Excited Electron–Ion Dynamics*, Comput. Phys. Commun. **151**, 60 (2003).

[70] A. Castro, H. Appel, M. Oliveira, C. A. Rozzi, X. Andrade, F. Lorenzen, M. A. L. Marques, E. K. U. Gross, and A. Rubio, *Octopus: A Tool for the Application of Time-Dependent Density Functional Theory*, Phys. Status Solidi **243**, 2465 (2006).

[71] X. Andrade, D. Strubbe, U. De Giovannini, A. H. Larsen, M. J. T. Oliveira, J. Alberdi-Rodriguez, A. Varas, I. Theophilou, N. Helbig, M. J. Verstraete, L. Stella, F. Nogueira, A. Aspuru-Guzik, A. Castro, M. A. L. Marques, and A. Rubio, *Real-Space Grids and the Octopus Code as Tools for the Development of New Simulation Approaches for Electronic Systems*, Phys. Chem. Chem. Phys. **17**, 31371 (2015).

[72] N. Tancogne-Dejean, M. J. T. Oliveira, X. Andrade, H. Appel, C. H. Borca, G. Le Breton, F. Buchholz, A. Castro, S. Corni, A. A. Correa, U. De Giovannini, A. Delgado, F. G. Eich, J. Flick, G. Gil, A. Gomez, N. Helbig, H. Hübener, R. Jestädt, J. Jornet-Somoza, A. H. Larsen, I. V Lebedeva, M. Lüders, M. A. L. Marques, S. T. Ohlmann, S. Pipolo, M. Rampp, C. A. Rozzi, D. A. Strubbe, S. A. Sato, C. Schäfer, I. Theophilou, A. Welden, and A. Rubio, *Octopus, a Computational Framework for Exploring Light-Driven Phenomena and Quantum Dynamics in Extended and Finite Systems*, J. Chem. Phys. **152**, 124119 (2020).

[73] M. A. L. Marques, C. A. Ullrich, F. Nogueira, A. Rubio, K. Burke, and E. K. U. Gross, *Time-Dependent Density Functional Theory*, in *Time-Dependent Density Functional Theory* (Springer, Berlin, Heidelberg, 2003).

[74] A. Scrinzi, *T-SURFF: Fully Differential Two-Electron Photo-Emission Spectra*, New J. Phys. **14**, 085008 (2012).

[75] U. De Giovannini, H. Hübener, and A. Rubio, *A First-Principles Time-Dependent Density Functional Theory Framework for Spin and Time-Resolved Angular-Resolved Photoelectron Spectroscopy in Periodic Systems*, J. Chem. Theory Comput. **13**, 265 (2017).

[76] T. E. Glover, R. W. Schoenlein, A. H. Chin, and C. V Shank, *Observation of Laser Assisted Photoelectric Effect and Femtosecond High Order Harmonic Radiation*, Phys. Rev. Lett. **76**, 2468 (1996).

[77] L. B. Madsen, *Strong-Field Approximation in Laser-Assisted Dynamics*, Am. J. Phys. **73**, 57 (2004).

[78] A. S. Kheifets, *The Attoclock and the Tunneling Time Debate*, J. Phys. B At. Mol. Opt. Phys. **53**, 72001 (2020).

[79] K. Amini, J. Biegert, F. Calegari, A. Chacón, M. F. Ciappina, A. Dauphin, D. K. Efimov, C. F. de M. Faria, K. Giergiel, P. Gniewek, A. Landsman, M. Lesiuk, M. Mandrysz, A. S. Maxwell, R. Moszynski, L. Ortmann, J. A. Perez-Hernandez,

A. Picon, E. Pisanty, J. S. Prauzner-Bechcicki, K. Sacha, N. Suárez, A. Zair, J. Zakrzewski, and M. Lewenstein, *Symphony on Strong Field Approximation*, Reports Prog. Phys. **82**, 116001 (2019).

[80] Z. Tao, C. Chen, T. Szilvási, M. Keller, M. Mavrikakis, H. Kapteyn, and M. Murnane, *Direct Time-Domain Observation of Attosecond Final-State Lifetimes in Photoemission from Solids*, Science **353**, 62 (2016).

[81] S. Fabian, N. Sergej, B. Peter, H. Matthias, S. Christian, F. Sebastian, T.-S. Miquel, S. V. M., K. E. E., K. N. M., F. Stephan, M. R. Díez, E. P. M., K. A. K., M. Norbert, P. Walter, and H. Ulrich, *Angular Momentum–Induced Delays in Solid-State Photoemission Enhanced by Intra-Atomic Interactions*, Science **357**, 1274 (2017).

[82] M. V. Berry, *Quantal Phase Factors Accompanying Adiabatic Changes*, Proc. R. Soc. London. A. Math. Phys. Sci. **392**, 45 (1984).

[83] D. Shin, S. A. Sato, H. Hübener, U. De Giovannini, J. Kim, N. Park, and A. Rubio, *Unraveling Materials Berry Curvature and Chern Numbers from Real-Time Evolution of Bloch States*, Proc. Natl. Acad. Sci. **116**, 4135 (2019).

[84] F. H. M. Faisal and J. Z. Kamiński, *Floquet-Bloch Theory of High-Harmonic Generation in Periodic Structures*, Phys. Rev. A **56**, 748 (1997).

[85] H. Hübener, M. A. Sentef, U. De Giovannini, A. F. Kemper, and A. Rubio, *Creating Stable Floquet-Weyl Semimetals by Laser-Driving of 3D Dirac Materials*, Nat. Commun. **8**, 13940 (2017).

[86] T. N. Ikeda, K. Chinzei, and H. Tsunetsugu, *Floquet-Theoretical Formulation and Analysis of High-Order Harmonic Generation in Solids*, Phys. Rev. A **98**, 63426 (2018).

[87] O. Neufeld, D. Podolsky, and O. Cohen, *Floquet Group Theory and Its Application to Selection Rules in Harmonic Generation*, Nat. Commun. **10**, 405 (2019).

[88] U. De Giovannini and H. Hübener, *Floquet Analysis of Excitations in Materials*, J. Phys. Mater. **3**, 12001 (2019).

[89] Y. H. Wang, H. Steinberg, P. Jarillo-Herrero, and N. Gedik, *Observation of Floquet-Bloch States on the Surface of a Topological Insulator*, Science **342**, 453 LP (2013).

[90] F. Mahmood, C.-K. Chan, Z. Alpichshev, D. Gardner, Y. Lee, P. A. Lee, and N. Gedik, *Selective Scattering between Floquet–Bloch and Volkov States in a Topological Insulator*, Nat. Phys. **12**, 306 (2016).

[91] S. Aeschlimann, S. A. Sato, R. Krause, M. Chávez-Cervantes, U. De Giovannini, H. Hübener, S. Forti, C. Coletti, K. Hanff, K. Rossnagel, A. Rubio, and I. Gierz, *Survival of Floquet–Bloch States in the Presence of Scattering*, Nano Lett. **21**, 5028 (2021).

[92] E. J. Sie, *Valley-Selective Optical Stark Effect in Monolayer WS2 BT - Coherent Light-Matter Interactions in Monolayer Transition-Metal Dichalcogenides*, in edited by E. J. Sie (Springer International Publishing, Cham, 2018), pp. 37–57.

[93] S. A. Sato, J. W. McIver, M. Nuske, P. Tang, G. Jotzu, B. Schulte, H. Hübener, U. De Giovannini, L. Mathey, M. A. Sentef, A. Cavalleri, and A. Rubio, *Microscopic Theory for the Light-Induced Anomalous Hall Effect in Graphene*, Phys. Rev. B **99**, 214302 (2019).

[94] J. W. McIver, B. Schulte, F.-U. Stein, T. Matsuyama, G. Jotzu, G. Meier, and A. Cavalleri, *Light-Induced Anomalous Hall Effect in Graphene*, Nat. Phys. **16**, 38 (2020).

[95] H. Hübener, U. De Giovannini, and A. Rubio, *Phonon Driven Floquet Matter*, Nano Lett. **18**, 1535 (2018).

[96] O. Neufeld and O. Cohen, *Optical Chirality in Nonlinear Optics: Application to High Harmonic Generation*, Phys. Rev. Lett. **120**, 133206 (2018).

[97] E. Pisanty, G. J. Machado, V. Vicuña-Hernández, A. Picón, A. Celi, J. P. Torres, and M. Lewenstein, *Knotting Fractional-Order Knots with the Polarization State of Light*, Nat. Photonics **13**, 569 (2019).

[98] E. Pisanty, L. Rego, J. San Román, A. Picón, K. M. Dorney, H. C. Kapteyn, M. M. Murnane, L. Plaja, M. Lewenstein, and C. Hernández-Garcia, *Conservation of Torus-Knot Angular Momentum in High-Order Harmonic Generation*, Phys. Rev. Lett. **122**, 203201 (2019).

[99] L. Rego, K. M. Dorney, N. J. Brooks, Q. L. Nguyen, C.-T. Liao, J. San Román, D. E. Couch, A. Liu, E. Pisanty, M. Lewenstein, L. Plaja, H. C. Kapteyn, M. M. Murnane, and C. Hernández-Garcia, *Generation of Extreme-Ultraviolet Beams with Time-Varying Orbital Angular Momentum*, Science **364**, eaaw9486 (2019).

[100] D. Ayuso, O. Neufeld, A. F. Ordonez, P. Decleva, G. Lerner, O. Cohen, M. Ivanov, and O. Smirnova, *Synthetic Chiral Light for Efficient Control of Chiral Light–Matter Interaction*, Nat. Photonics **13**, 866 (2019).

[101] Gavriel Lerner, O. Neufeld, L. Hareli, G. Shoulga, E. Bordo, A. Fleischer, D. Podolsky, A. Bahabad, and O. Cohen, *Multi-Scale Dynamical Symmetries and Selection Rules in Nonlinear Optics*, ArXiv:2109.01941 (2021).

[102] M. Even Tzur, O. Neufeld, A. Fleischer, and O. Cohen, *Selection Rules for Breaking Selection Rules*, New J. Phys. **23**, 103039 (2021).

[103] M. E. Tzur, O. Neufeld, E. Bordo, A. Fleischer, and O. Cohen, *Selection Rules in Symmetry-Broken Systems by Symmetries in Synthetic Dimensions*, Nat. Commun. **13**, 1312 (2022).

[104] C. Hartwigsen, S. Goedecker, and J. Hutter, *Relativistic Separable Dual-Space Gaussian Pseudopotentials from H to Rn*, Phys. Rev. B **58**, 3641 (1998).

[105] O. Neufeld and O. Cohen, *Background-Free Measurement of Ring Currents by Symmetry-Breaking High-Harmonic Spectroscopy*, Phys. Rev. Lett. **123**, 103202 (2019).